\documentclass[runningheads]{llncs}
\usepackage{graphicx}

\usepackage[english]{babel}
\usepackage[utf8]{inputenc}

\usepackage[  square,        %
comma,         %
numbers,       %
sort&compress, %
]{natbib}
\usepackage[cmex10]{amsmath}
\usepackage{amsfonts,amssymb}
\usepackage{mathtools}
\usepackage{bm}
\usepackage{xspace}
\usepackage{booktabs}
\usepackage{multirow}
\usepackage{microtype}
\usepackage{xcolor}
\usepackage{paralist}
\usepackage{enumitem} %

\usepackage{tikz}
\usepackage{pgfplots}
\pgfplotsset{compat=1.8}
\usetikzlibrary{pgfplots.statistics,patterns}
\usetikzlibrary{calc} %
\pgfplotsset{compat=newest}
\usetikzlibrary{positioning}
\usetikzlibrary{arrows, matrix}
\usetikzlibrary{fit}
\usepgfplotslibrary{groupplots}

\definecolor{goeblue}{RGB}{0,51,102}

\renewcommand{\paragraph}[1]{{\vskip 8pt \noindent\it #1. }}

\usetikzlibrary{external}

\usepackage[caption=false,font=footnotesize]{subfig}

\usepackage{url}

\pgfplotsset{
	coline1/.style={color=black},
	coline2/.style={color=red, densely dashed},
	coline3/.style={color=red!50!black, densely dotted},
	coline4/.style={color=green!50!black, densely dashdotted},
	coline5/.style={color=purple, densely dashdotdotted},
	coline6/.style={color=gray!50!black, dotted},
	coline7/.style={color=blue!50!black, loosely dashdotdotted},
}
\tikzset{
	>=stealth',
	colineA1/.style={color=black},
	colineA2/.style={color=red, densely dashed},
	colineA3/.style={color=red!50!black, densely dotted},
	colineA4/.style={color=green!50!black, densely dashdotted},
	colineA5/.style={color=purple, densely dashdotdotted},
	colineA6/.style={color=gray!50!black, dotted},
	colineA7/.style={color=blue!50!black, loosely dashdotdotted},
}

\newcommand{\DCT}{\ensuremath{\bm{D}}\xspace}

\newcommand{\x}{\ensuremath{\bm{X}}\xspace}
\newcommand{\y}{\ensuremath{\bm{Y}}\xspace}

\newcommand{\prnu}{\ensuremath{\bm{K}}\xspace}

\newcommand{\w}{\ensuremath{\bm{W}}\xspace}
\newcommand{\I}{\ensuremath{\bm{I}}\xspace}
\newcommand{\J}{\ensuremath{\bm{J}}\xspace}
\newcommand{\R}{\ensuremath{\bm{X}}\xspace}
\newcommand{\Z}{\ensuremath{\bm{\tilde{X}}}\xspace}
\newcommand{\Ry}{\ensuremath{\bm{Y}}\xspace}
\newcommand{\RyZ}{\ensuremath{Y}\xspace}
\newcommand{\Zy}{\ensuremath{\bm{\tilde{Y}}}\xspace}
\newcommand{\ZyZ}{\ensuremath{\tilde{Y}}\xspace}

\newcommand{\qqPo}{\ensuremath{\rho\xspace}}
\newcommand{\qqSam}{\ensuremath{r\xspace}}
\newcommand{\qqMLSame}{\ensuremath{\phi_1\xspace}}
\newcommand{\qqMLDiff}{\ensuremath{\phi_2\xspace}}

\newcommand{\ma}{\ensuremath{\bullet}} %

\newcommand\independent{\protect\mathpalette{\protect\independenT}{\perp}}
\def\independenT#1#2{\mathrel{\rlap{$#1#2$}\mkern2mu{#1#2}}}

\begin{document}

\title{On the Security and Applicability of\\Fragile 
	Camera Fingerprints\thanks{Published at ESORICS 2019}}
\titlerunning{Security and Applicability of Fragile 
	Camera Fingerprints}
\author{Erwin Quiring\inst{1} \and
Matthias Kirchner\inst{2} \and
Konrad Rieck\inst{1}}
\authorrunning{E. Quiring et al.}
\institute{TU Braunschweig, Germany \and
Binghamton University, USA
}
\maketitle              %

\vspace{-.51cm}

\begin{abstract}
	
Camera sensor noise is one of the most reliable device characteristics 
in digital image forensics, enabling the unique linkage of images to 
digital cameras. This so-called camera fingerprint gives rise
to different applications, such as image forensics and authentication.
However, if images are publicly available, an adversary can 
estimate the fingerprint from her victim and plant it into spurious 
images.
The concept of fragile camera fingerprints addresses this attack
by exploiting asymmetries in data access: While the camera owner 
will always have access to a full fingerprint from uncompressed images, 
the adversary has typically access to compressed images and thus only 
to a truncated fingerprint.
The security of this defense, however, has not been systematically 
explored yet.
This paper provides the first comprehensive 
analysis of fragile camera fingerprints under attack. 
A series of theoretical and practical tests demonstrate that fragile 
camera fingerprints allow a reliable device identification for common 
compression levels in an adversarial environment.

\end{abstract}
\section{Introduction}\label{sec:Intro}

Minimal, inevitable manufacturing imperfections of digital camera 
sensors lead to the photo-response non-uniformity (PRNU) signal, a
highly unique and reliably detectable camera device characteristic%
~\cite{Fridrich:2013dq}. Similar to a robust digital watermark, the 
PRNU signal is unnoticeably present in any image taken by the same 
camera, but differs between images from different cameras. 
These properties make the PRNU a natural \emph{camera fingerprint}. 
It has found widespread applications in forensics to attribute digital 
images to their source camera~\cite{Fridrich:2013dq}. Recent works have 
also proposed to use the PRNU as a means to link mobile device 
authentication to inherent hardware characteristics of the 
mobile device~\cite{ValColBia+17,BaFuKou+18}.

In practice, however, these use cases face the problem of 
\emph{fingerprint copy-attacks}~\cite{Lukas:2006aa,Gloe:2007ab}. 
If Alice shares images from her camera with the public, Mallory can 
estimate Alice's fingerprint, plant it into her images, and pretend 
that an arbitrary image was captured by Alice's camera.
The so-called \emph{triangle test} \cite{Goljan:2011aa} detects such 
attacks \emph{ex post}, but it potentially requires an exhaustive 
search over all public images shared by Alice. A proactive defense 
based on the 
notion of \emph{fragile camera fingerprints} has recently been proposed 
by Quiring and Kirchner \cite{QuiringKirchner2015} for scenarios that 
warrant camera identification from high-quality (uncompressed) images. 
Here, the camera owner Alice can exploit an asymmetry in the quality of 
accessible data by only sharing JPEG-compressed images with the public 
while retaining her uncompressed images private. As a result, she will 
always be able to provide her full fingerprint from high-quality images 
when asked to do so. In contrast, Mallory's estimate of Alice's fingerprint 
from public JPEG images will only contain the part that is robust to 
lossy JPEG compression while lacking the fragile component. A test for 
the presence of the fragile fingerprint will then prevent Mallory from 
making an uncompressed image look like one of Alice's uncompressed 
images.

In forensics applications, fragile camera fingerprints are of particular 
relevance to the prevention of fingerprint-copy attacks in support of 
high-quality image forgeries, which may otherwise convey a false sense 
of trustworthiness~\cite{Bohme-IWCF2009}. Equally important, fragile 
fingerprints are currently the only scalable approach to establish 
mobile device authentication based on physical camera characteristics 
that mitigates fingerprint leakage from 
public images: conducting the triangle test \cite{Goljan:2011aa} on 
every authentication attempt is computationally infeasible, and an 
alternative proposal for a targeted fingerprint-copy attack detector 
by Ba et al.~\cite{BaFuKou+18} can be defeated by an adversary with 
two cameras.

The practical applicability of fragile camera fingerprints in these 
security-related scenarios crucially depends on their robustness 
against attacks. As Quiring and Kirchner's work 
\cite{QuiringKirchner2015} only 
provided preliminary results in this regard, this paper sets out 
to deliver a thorough and more comprehensive security analysis. 
Specifically, we examine the amount of information that Mallory can 
estimate, recover 
and exploit in a series of theoretical and empirical considerations. First, 
we analytically derive an upper bound on the correlation between 
Alice's and Mallory's fingerprint estimates with respect to the JPEG 
quality of 
publicly shared images Mallory has access to. Second, to test for 
dependencies beyond linear correlation, a kernel statistical test is 
used to assess whether Alice's fragile fingerprint is statistically 
independent 
of Mallory's fingerprint. Third, we demonstrate that practical attempts 
to recover quantized JPEG coefficients from potentially remaining 
dependencies do not increase Mallory's ability to mount successful 
attacks. Fourth, we test the resistance of fragile fingerprints against 
practical fingerprint-copy attacks. We finally illustrate that fragile 
fingerprints and the triangle test are a powerful combination in 
forensics applications.

The rest of this paper is organized as follows. Section~\ref{sec:RelatedWork} reviews the background of sensor noise forensics before Section~\ref{sec:OurMethod} discusses fragile fingerprints and their possible applications. Section~\ref{sec:sec_analysis} 
provides a comprehensive analysis of Mallory's attack surface, while 
Section~\ref{sec:Application} reports on experiments around the 
applicability of fragile fingerprints. Section~\ref{sec:conclusion} 
concludes the paper.

\section{Background} \label{sec:RelatedWork}

Before introducing fragile fingerprints, we give a short primer on 
camera identification, the possible fingerprint-copy attack, and the 
triangle test as defense.
Throughout our work, the \emph{notation} is as follows: vectors and 
matrices are set in boldface font. Operations on vectors and matrices 
are point-wise if not stated otherwise; the operator $\ma$ denotes 
matrix multiplication.

\subsection{Camera Identification from Sensor Noise Fingerprints}
\label{subsec:fingerprint-estimate}

Due to sensor element manufacturing imperfections, each camera image 
does not only contain the original noise-free image content $\I^0$, but 
also the PRNU \prnu as a camera-specific, multiplicative noise 
factor. A common simplified model of the image capturing process assumes
the final image $\I$ to take the form \cite{Fridrich:2013dq}
\begin{align}
\I = \I^0 + \I^0 \prnu + \bm{\Gamma} \, ,
\end{align}
where $\bm{\Gamma}$ reflects a variety of other additive noise terms. 
Due to its multiplicative nature, the PRNU is not present 
in images with dark scene contents (i.\,e.,~$\I^0 \approx \bm{0}$).
Extensive experiments have demonstrated that the PRNU factor \prnu 
represents a unique and robust camera fingerprint \cite{Goljan:2009aa} 
that can be estimated from a number of images $\I_1,\dots,\I_N$ 
taken with a given camera of interest. The standard approach utilizes 
a denoising filter $F(\cdot)$ and models noise residuals 
$\w_k  = \I_k - F(\I_k)$ as \cite{Fridrich:2013dq}
\begin{align}
\w_k =  \I_k \prnu + \bm{\Theta}_k\, .
\label{eq:noiseResidual}
\end{align}
Modeling noise $\bm{\Theta}$ subsumes $\bm{\Gamma}$ and residues of 
the image content due to inherent~imperfections of the denoising filter 
in separating image content from noise. 
Adopting an i.i.d.~Gaussian noise assumption for $\bm{\Theta}$,
the maximum likelihood (ML) estimator of \prnu is \cite{Fridrich:2013dq}
\begin{align}
\hat{\prnu} = \left(\kern1pt \sum_{k=1}^N \w_k \I_k \right) \cdot 
\left(\kern1pt \sum_{k=1}^N (\I_k)^2 \right)^{-1} \, . \label{eq:prnu}
\end{align}
A more simple estimator takes the pixel-wise average of the noise 
residuals \cite{Lukas:2006aa}. A post-processing step is recommended to 
clean $\hat{\prnu}$ from so-called non-unique artifacts, e.\,g., due to 
demosaicing or lens distortion correction 
\cite{Fridrich:2013dq,Goljan:2012aa,Gloe:2012pt}. Given a query image 
\J of unknown provenance, camera identification then works by computing 
the residual $\w_{\J} = \J - F(\J)$, and evaluating its similarity to a 
camera fingerprint estimate against a set threshold $\tau$,
\begin{align}
\phi_{\w_{\J},\J\hat{\prnu}} = \mathrm{sim}( \w_{\J} , \J \hat{\prnu} 
)\, \gtrless \tau.
\label{eq:camera-identification}
\end{align}
Suitable similarity measures for this task are normalized correlation 
or peak-to-correlation energy (PCE) 
\cite{Lukas:2006aa,Fridrich:2013dq}.

\subsection{Fingerprint-Copy 
	Attack}\label{subsec:fingerprint-copy-method}

Following the procedure described in 
Section~\ref{subsec:fingerprint-estimate}, Mallory may obtain an 
estimate 
of Alice's camera fingerprint from a set of $N_E$ publicly available 
images. Denoting this estimate $\hat{\prnu}_E$, Mallory can then 
attempt to 
make an arbitrary image \J look as if it was captured by Alice's 
camera. The multiplicative nature of PRNU suggests a \emph{fingerprint 
copy attack} of the form \cite{Lukas:2006aa}
\begin{align}
\J' = [\J (1+\alpha \hat{\prnu}_E)]\, ,
\label{eq:fingerprint-copy-attack}
\end{align}
with $\alpha>0$ being the scalar fingerprint strength parameter. 
Attacks of this type have been demonstrated to be effective, in the 
sense that they can successfully mislead a camera identification 
algorithm in the form of Equation~\eqref{eq:camera-identification}. 
However, the attack's success generally depends on a good choice 
of~$\alpha$: too low values mean that the bogus image $\J'$ may not~be 
assigned to Alice's camera; a too strong embedding will make the image 
appear suspicious \cite{Goljan:2011aa,Marra:2014aa}. In practical 
scenarios, Mallory may have to apply further processing to make 
her 
forgery more compelling, e.\,g., removing the genuine camera 
fingerprint \cite{Karakucuk:2015aa,Entrieri:2016aa}, synthesizing 
demosaicing artifacts \cite{Kirchner:2009aa}, and removing or adding 
traces of JPEG compression~\cite{Stamm:2011aa}.

\subsection{Triangle Test}\label{subsec:TriangleTest}

Under realistic assumptions, it is impossible to prevent Mallory from 
forcing a high similarity score in 
Equation~\eqref{eq:camera-identification} for arbitrary images from a 
foreign camera. Yet Alice can utilize that noise residuals computed 
with practical denoising filters will always contain remnants of image 
content to establish that image  $\J'$ underwent a fingerprint-copy 
attack \cite{Goljan:2011aa}. The key observation here is that the 
already existing similarity  between a noise residual $\w_{\I}$ from an 
image \I taken with Alice's camera and the noise residual $\w_{\J'}$ 
due to a common PRNU term will be slightly increased  by some shared 
residual image content, if \I contributed to Mallory's fingerprint 
estimate 
$\hat{\prnu}_E$. Alice can thus test which of her public images have 
been used by Mallory to mount the attack by evaluating whether the 
similarity of their noise residuals $\w_{\I}$ with $\w_{\J'}$ is 
suspiciously large.

Because the additional correlation imposed by shared image~content is 
generally rather weak and also varies with~macroscopic image 
characteristics, Goljan et al.~\cite{Goljan:2011aa} propose a 
\emph{triangle test} to calibrate the test statistic. Specifically, the 
test does not only consider the observed correlation 
$\nu_{\w_{\I},\w_{\J'}}$ between residuals $\w_{\I}$ and 
$\w_{\J'}$, but 
it also employs a correlation predictor to estimate the correlation 
$\tilde{\nu}_{\w_{\I},\w_{\J'}}$ between $\w_{\I}$ and $\w_{\J'}$ if 
image \I had \emph{not} participated in the computation of 
$\hat{\prnu}_E$. This predictor takes the correlation between Alice's 
own fingerprint and both $\w_{\I}$ and $\w_{\J'}$ into account---hence 
the name triangle test. Assuming a linear relationship between the 
observed and the predicted correlation, the proposed test then 
evaluates 
\begin{align}
\nu_{\w_{\I},\w_{\J'}} - \theta \kern1pt \tilde{\nu}_{\w_{\I},\w_{\J'}} 
- \mu \gtrless t 
\label{eq:triangle-test}
\end{align}
for a suitably chosen threshold $t$. The parameters $\theta$ and $\mu$ 
are estimated %
from a set of \emph{safe} images, for 
which it can be guaranteed that they have not been used by Mallory. 
We refer to \citet{Goljan:2011aa} for a detailed exposition of the 
correlation predictor and the parameter estimators. The test statistic 
in Equation~\eqref{eq:triangle-test} is expected to have zero mean when 
two noise residuals share only a common PRNU term. 
A larger difference indicates an additional shared term, possibly due 
to image \I's involvement in the attack. 
Observe that Alice may have to test \emph{all images} ever made public 
by her as part of a comprehensive defense. We finally point out that a 
number of fingerprint-copy attack variations have been proposed 
recently that are reportedly less likely to be exposed by the triangle 
test~\cite[e.g.][]{Marra:2014aa}.

\section{Fragile Camera Fingerprint} \label{sec:OurMethod}
As a novel and proactive defense against fingerprint-copy attacks,
Quiring and Kirchner~\cite{QuiringKirchner2015} introduce the 
notion of fragile camera fingerprints that vanish under lossy JPEG 
compression.
The idea is based on two mild assumptions: 1)~Alice's device 
supports capturing images in uncompressed format, which is true 
nowadays for many devices operating under mobile platforms, such as iOS 
and Android; 
2) Alice only shares JPEG images with the public, which is already today's 
quasi-standard for image online storage and sharing.
When combined, these two assumptions allow Alice to effectively exploit an 
asymmetry in the quality of data access. With full access to her camera, Alice 
is always in the position to present a fingerprint estimate $\hat{\prnu}$ from uncompressed images while Mallory is restricted to estimate 
$\hat{\prnu}_E$ from JPEG-compressed images.

On a technical level, the concept of fragile camera fingerprints exploits the lossy nature of JPEG compression. JPEG maps each non-overlapping $8\times 8$ pixel block in an image to $8\times 8$  discrete cosine transform (DCT) coefficients. Each of the 64 coefficients quantifies the influence of a particular frequency subband and will be quantized based on an $8\times 8$ quantization table with quantization factors for the 64 DCT subbands. Larger quantization factors mean that the DCT coefficients in the corresponding subband are more likely to be quantized to zero. Quantization factors generally increase with decreasing JPEG quality and grow towards the bottom right corner of the quantization table to suppress high-frequency image details more aggressively.

In consequence, Mallory's camera fingerprint estimate from JPEG-compressed 
images will be strongly distorted in the high-frequency DCT subbands 
due to larger quantization errors. If the quantization is too strong, 
Mallory's images will lack high-frequency content altogether and so 
will her fingerprint estimate. In other words, her estimate only 
comprises the fingerprint component that is robust to JPEG compression. 
A fingerprint estimate from uncompressed images is in 
turn distributed almost evenly over all subbands \cite{QuiringKirchner2015}.
Hence, Alice has access to a \emph{fragile camera fingerprint}, 
computed from the high-frequency subbands~only.

\begin{figure}[t]
	\centering
	\includegraphics[]{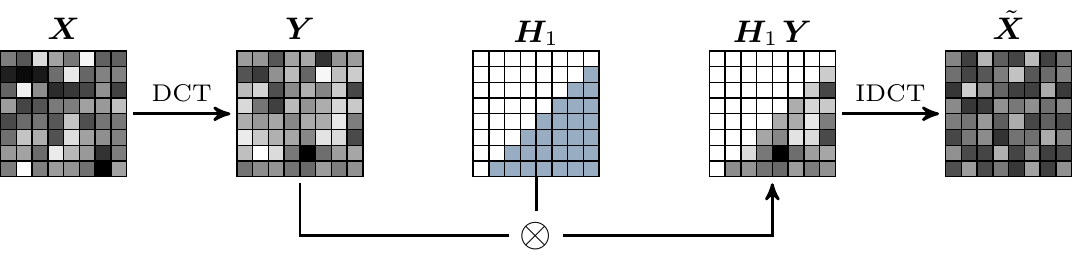}
	\vspace{-0.7em}
	\caption{Fragile fingerprint computation based on a 
	subband-selective filter~$H_1(\x)$: Each pixel block \x is 
	mapped to its DCT 
		representation \y, element-wise multiplied by a binary mask 
		$\bm{H}_1$, and transformed back to the spatial domain to give $\tilde\x$.}
	\label{fig:highPassFilter}
\end{figure}

To obtain the fragile part, it is instructive to define a mode-selective 
highpass filter $H_c(\cdot)$. Based on a binary multiplicative mask 
$\bm{H}_c = [h_{i,j}], 1 \leqslant i,j \leqslant 8$, the filter retains 
a defined set of DCT subbands and sets all other subband coefficients 
to zero. Alice should choose $\bm{H}_c$ depending on the maximum JPEG 
quality of her published images. For a sufficiently conservative 
choice, she can assume that the retained subbands are available 
exclusively to her.
Quiring and Kirchner \cite{QuiringKirchner2015} propose to parameterize 
$\bm{H}_c$ with a cut-off along the $(-c)$-th anti-diagonal of the 
DCT coefficient matrix,
\begin{align}
h_{i,j} = [(i + j - 8 - c) > 0] \; , \label{eq:filter}
\end{align}
where $[\cdot]$ denotes the Iverson bracket.
Figure~\ref{fig:highPassFilter} summarizes the internal steps of 
$H_c(\cdot)$ for cut-off parameter $c=1$, which retains all DCT 
subbands in the lower right triangle.
Equipped with $H_c$, a refined similarity test of the form
\begin{align}
\phi_{\w_{\J},\J\hat{\prnu}}(c) = \mathrm{sim}\left( H_c(\w_{\J}), H_c(\J \hat{\prnu}) \right) \gtrless \tau  \, .
\label{eq:SimiliarityMeasureHighFreq}
\end{align}
then establishes camera identification from fragile fingerprints, which is of particular relevance in the following two application scenarios.

\paragraph{Digital Image Forensics}
Testing for the presence of camera fingerprints facilitates device 
identification and image manipulation detection in forensic 
applications~\cite{Fridrich:2013dq}. Fragile camera fingerprints can 
benefit scenarios that warrant the analysis of high-quality images, for 
instance when uncompressed images ought to be presented as a source of 
particularly high trustworthiness. In this case, Alice can establish 
that a spurious image was not captured by her camera by having kept her 
uncompressed images private. She presents the fragile fingerprint when 
needed, potentially in combination with cryptographic 
safeguards~\cite{Mohanty+19}  
or in some form of zero-knowledge proof to further secure her fragile 
fingerprint from leakage.

\paragraph{Mobile Device Authentication}
Camera fingerprints have been proposed as building blocks for 
augmenting mobile device authentication schemes with physical hardware 
characteristics~\cite{Bojinov+14,ValColBia+17,BaFuKou+18}.
Yet like with device signatures from other types of 
hardware sensors, the vulnerability to fingerprint-copy\,/\,spoofing 
attacks is of particular concern \cite{Alaca+16}. To the best of our 
knowledge, the concept of fragile fingerprints is the only existing 
approach that would address the problem in a proactive and scalable 
manner. Performing a triangle test upon every authentication attempt is 
computationally infeasible, let alone that Alice may object to the idea 
of sharing all her images with the service she wants to authenticate 
to.

Ba et al.~\cite{BaFuKou+18} attempt to work around that issue in 
their authentication protocol by requiring the user to take two images 
of different visual codes during the 
identification phase. Similar to the triangle test, the reasoning is 
that two spoofed images will correlate implausibly strongly as they do 
not only share Alice's fingerprint but also the one from Mallory's 
camera. This measure can be easily circumvented by an attacker who uses 
two different devices to take the respective pictures however. In 
this way, Mallory prevents the additionally shared signal.

\vspace{4pt}
\noindent In summary, fragile camera fingerprints enable novel applications in 
image forensics and mobile device authentication. Their applicability, 
however, depends on the robustness against attacks. In the remainder of 
this work, we thus perform a comprehensive security analysis.

\section{Security Analysis}\label{sec:sec_analysis}

A secure application of fragile camera fingerprints demands that 
Mallory cannot estimate the fragile fingerprint from JPEG-compressed 
images.
We guide our analysis along the following three questions:
\begin{description}[before={\renewcommand\makelabel[1]{\normalfont 
##1}},leftmargin=1.825\parindent]
	\setlength\itemsep{0.25em}
	\item[\textit{(Q1)}] Can we bound the quality of 
	Mallory's fingerprint estimate~$\bm{\hat{K}_E}$?
	\item[\textit{(Q2)}] Can Mallory improve her fingerprint estimate 
	by exploiting the quantized high-frequency or robust 
	low-frequency information?
	\item[\textit{(Q3)}] Can Mallory perform a successful 
	fingerprint-copy attack?
\end{description}

\subsection{Datasets and Experimental Setup}\label{subsec:dataset}

Where empirical tests are warranted, we adopt the setup described
by Quiring and Kirchner~\cite{QuiringKirchner2015}. 
The dataset consists of images from the 
Dresden~Image~Database \cite{Gloe:2010ad} (DDB) %
and the RAISE Image Database~\cite{Dang-Nguyen:2015aa}, 
cf.~Table~\ref{tab:imageset}. 
\begin{table}[b]
	\scriptsize
	\setlength{\tabcolsep}{6.75pt}
	\caption{Number of images per test set and camera.}
	\label{tab:imageset}
	\centering
	\begin{tabular}{llcccc}
		\toprule
		\multirow{2}{*}{Database} & \multirow{2}{*}{Camera model} & 
		\multicolumn{2}{c}{Fingerprint estimate} & 
		\multicolumn{2}{c}{Benchmark data} \\
		\cmidrule(lr){3-4} \cmidrule(lr){5-6}
		&  &  Camera 0 & Camera 1 & Camera 0 & Camera 1 \\ 
		\midrule
		\multirow{3}{*}{Dresden~\cite{Gloe:2010ad}} & 
		Nikon D70 & 25 & 25 & 175  & 188  \\
		& Nikon D70s & 25 & 25 & 175  & 174  \\
		& Nikon D200 & 25 & 25 & 360  &  370 \\
		\cmidrule(lr){2-6}
		RAISE~\cite{Dang-Nguyen:2015aa} & Nikon D7000 & 300 & --- & 
		4648 & --- \\
		\bottomrule
	\end{tabular}
\end{table}
In particular,
we use 25 homogeneously lit flat field images of each DDB
camera to obtain uncompressed fingerprint estimates. 1442 
natural images serve as benchmark data. We present aggregated results 
over the six cameras in the following, as all gave similar results.
The RAISE database only provides natural images. We randomly select 300 
images for fingerprint computation, leaving us with 4648 images for a
benchmark set that facilitates the study of attacks where Mallory has 
access to a large number of public images. 
Note that the usage of 300~natural images for fingerprint estimation 
can be attributed to the heterogenous content of natural images. 
In an authentication scenario, a user can be asked
to take a much smaller number of suitable images 
(e.\,g. a white wall) without nuisance image content.
If not stated otherwise, we use 
the standard Wavelet denoiser to obtain noise residuals \cite{Lukas:2006aa} and 
the ML formulation in 
Equation~\eqref{eq:prnu} to estimate fingerprints. 

In order to also guide our evaluation at the practically used 
JPEG quality, we collected over 1.4 Million JPEG images from 
Twitter, Instagram, Imgur, Deviantart and Flickr. The average JPEG 
quality is 83.5 with a standard deviation of 9.2.
This, for instance, fits to recommendations from Flickr, Wikimedia and 
the official Android documentation that recommend qualities less 
than~90.

\subsection{(Q1) Analytical Quality of Fingerprint Estimation} 
\label{subsec:theoModel_fingestimation}

Our first objective is to establish a bound on the quality of Mallory's 
fingerprint estimate from Alice's camera irrespective of a concrete 
image data set to reflect Mallory's chances of performing a 
successful fingerprint-copy attack.
In particular, we adopt the notion of \emph{quality of fingerprint 
estimation} by Goljan et al.~\cite{Goljan:2011aa}. We derive an 
analytical expression for the correlation between Alice's fingerprint 
from uncompressed images, $\bm{\hat{K}}$, and 
Mallory's fingerprint from compressed images,~$\bm{\hat{K}_E}$,
\begin{align}
\mathrm{cor}( \, H_c(\bm{\hat{K}}), \, H_c(\bm{\hat{K}_E}) 
\, ) \, . \label{eq:corFingerprintQualityTheo}
\end{align}
This quantity can be seen as a simplified version of the similarity 
measure in Eq.\,\eqref{eq:SimiliarityMeasureHighFreq} for images  
taken under ideal conditions, e.g.\ homogeneously lit. 
As we focus on high-frequency subbands only, less image content 
disturbs the fingerprint calculation. Thus,
$H_c(\J \hat{\prnu})$ resembles $H_c(\bm{\hat{K}})$, as well
$H_c(\w_{\J}) \approx H_c(\bm{\hat{K}_E})$.

We make three assumptions to simplify the calculation. 
First, Mallory computes her fingerprint from the same, but 
compressed, image set that Alice uses for her uncompressed 
estimate. This will yield a loose upper bound for cases where 
Mallory obtains a different JPEG-compressed image set that Alice has 
not used.
Moreover, we apply a simple fingerprint estimator that takes the 
pixel-wise average of noise residuals. Finally, we assume a negligible 
correlation between individual DCT subbands and across images.
The first and the second assumption imply that the fingerprint 
calculation can be modeled as pixel-wise averaging. 
Denote $\R_i$ the $i$-th uncompressed image from Alice's 
camera and $\Z_i$ its JPEG-compressed version to rewrite
Equation~\eqref{eq:corFingerprintQualityTheo}~as
\begin{align}
\mathrm{cor}( \, H_c(\bm{\hat{K}}), \, H_c(\bm{\hat{K}_E}) 
\, ) =
\mathrm{cor}\left(  \textstyle \sum_i H_c ( \R_i ) , \sum_i H_c ( 
\Z_i )\right) .
\label{eq:RhoTheoModel}
\end{align}
Appendix~\ref{sec:AppCorrInDCT} establishes that the 
\emph{sample correlation coefficient}
based on Equation~\eqref{eq:RhoTheoModel}, $r(c)$, can be computed 
in the DCT domain directly.
We write $\Ry_i$ and $\Zy_i$ for the DCT representations of $\R_i$ and 
$\Z_i$, respectively, to obtain
\begin{align}
\qqSam(c) \cong
\mathrm{cor}\left( \textstyle \sum_i (\bm{H}_c \; \Ry_i)  , \sum_i 
(\bm{H}_c \; \Zy_i) \right) \, .
\label{eq:RhoTheoModelDCT}
\end{align}
The coefficient is parametrized by the cut-off $c$ from 
Equation~\eqref{eq:filter}.
High-pass filter $H_c$ is now made explicit 
through the DCT mask $\bm{H}_c$ (see Section~\ref{sec:OurMethod}),
yielding a convenient formulation to compute the 
sample correlation coefficient
between Alice's and Mallory's fingerprint directly in the DCT domain.
This formulation thus allows us to use 
known statistical distribution models for~DCT~coefficients.

We continue to derive the \emph{population correlation coefficient} by assuming a Laplacian distribution for the AC DCT coefficients~\cite{LaplaceDistr}. 
Specifically, denote $\RyZ_{i,s}$ the random variable representing the $s$-th subband of  the $i$-the uncompressed image. Equivalently, denote $\ZyZ_{i,s}$ the respective quantized counterpart to reflect the effect of JPEG compression on $\RyZ_{i,s}$. 
Appendix~\ref{sec:AppQuantization} establishes the general relation 
between the two random variables in terms of their covariance, which 
can be expressed solely on the basis of the distribution of the 
uncompressed variable $\RyZ_{i,s}$. We highlight this by defining
$\mathrm{Cov^+}(\RyZ_{i,s}) = \mathrm{Cov}(\RyZ_{i,s}, \ZyZ_{i,s})$. 
A similar derivation for the variance yields 
$\mathrm{Var^+}(\RyZ_{i,s}) = \mathrm{Var}(\ZyZ_{i,s} )$.
Appendix~\ref{sec:AppQuantization} shows how aggregating these 
quantities over the various subbands $s\in S_c$ as specified by filter 
$H_c$ of various images leads to the following formulation for the 
population correlation coefficient:
\begin{align}
\qqPo(c) \cong \dfrac{\sum_i \sum_{s \in S_c} 
	\mathrm{Cov^+}(\RyZ_{i,s})}{\sqrt{\strut\sum_i\sum_{s \in S_c}  
		\mathrm{Var}(\RyZ_{i,s})}\sqrt{\strut\sum_i 
		\sum_{s \in S_c} 
		\mathrm{Var^+}(\RyZ_{i,s})}}   \label{eq:PopCorCoeff}\, .
\end{align}
This equation is a first step towards an analytical understanding of the impact of JPEG-induced quantization on the ability to estimate fragile camera fingerprints. 
Specifically, it allows Alice to deduce the expected 
correlation of Mallory's fingerprint with her fingerprint based on 
general DCT distribution assumptions. %
Note that the derived correlation is computed under the assumption of a strong attacker: 
Mallory bases her fingerprint estimation on the same image set as 
Alice; her images only differ in that they are JPEG-compressed.
Consequently, Equation~\eqref{eq:PopCorCoeff} can serve as upper bound 
for the more realistic scenario when Mallory has only access to a 
different image set. Alice, as camera owner, will always be 
able to create new images for her fingerprint. 
The next section demonstrates the validity of our analytical derivation 
under practical conditions for both~scenarios.

\subsection{(Q1) Empirical Quality of Fingerprint Estimation}
\label{subsec:evalTheoModel}

We start with the quality of Mallory's fingerprint estimate when 
Alice and Mallory operate on the same image set, and then continue with 
different image sets.

\begin{figure*}[t]
	\centering
	\captionsetup[subfigure]{labelformat=empty}
	\subfloat[]{\label{subfig:resA1} 
	\includegraphics[]{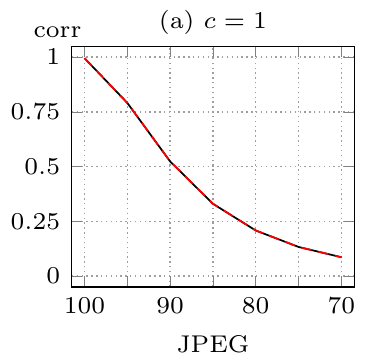}
	}
		 \hspace{.0em}
	\subfloat[]{\label{subfig:res1} 
	\includegraphics[]{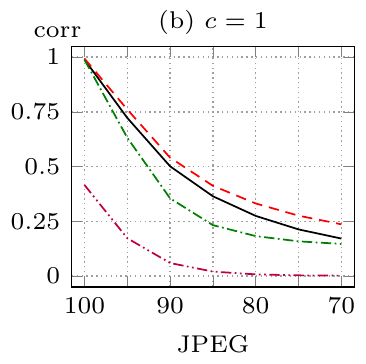}
	}
		 \hspace{.00em}
	\subfloat[]{\label{subfig:res2} 
	\includegraphics[]{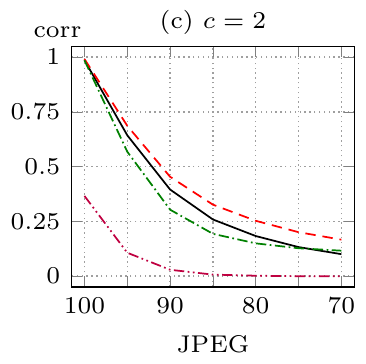}
	}
	\vspace{-2.5em}
	\subfloat[]{
		\label{subfig:resALegend}
\includegraphics[]{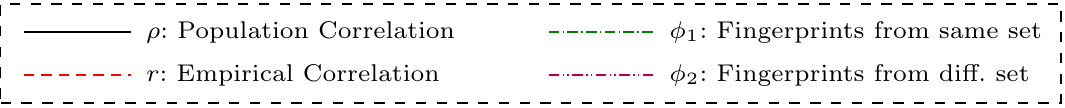}
	}
	\vspace{-1.0em}
	\caption{Quality of fingerprint estimation. Results from (a) 
	$N_E=250$ synthetic images and (b)-(c) $N_E=250$ natural Nikon 
	D7000 images.}
	\label{fig:resA}
\end{figure*}

\paragraph{Same image sets}
In a first step, we compute the population correlation 
coefficient~$\qqPo$ from Equation~\eqref{eq:PopCorCoeff} on a set of 
250~synthetic images. Each image follows a zero-mean Laplacian 
distribution with a randomly generated scale parameter.
This allows us to examine $\qqPo$ on idealized conditions. 
We compare $\qqPo$ with its empirical counterpart, $\qqSam$, 
as given in Equation~\eqref{eq:RhoTheoModelDCT}.
Figure~\ref{subfig:resA1} shows that the two derived quantities are 
consistent under varying JPEG compression levels.

In the next experiment, we use natural images from 
the Nikon D7000. %
We give Mallory access to \mbox{$N_E=250$} JPEG-compressed images, 
derived from the same set that Alice uses for her fingerprint.
Varying the JPEG quality and cut-off parameter~$c$, we compute $\qqPo$ 
and $\qqSam$.
The computation of $\qqPo$ involved a standard maximum likelihood estimator to obtain the Laplace scale parameter for each DCT subband per image. 
For benchmark purposes, we also include the
sample correlation coefficient $\qqMLSame$ between Alice's and 
Mallory's fingerprint, both calculated with the ML formulation 
in~Equation~\eqref{eq:prnu}. %
We repeat the experiments five 
times and report averaged results for $\qqPo, \qqSam$ and $\qqMLSame$ 
in Figures~\ref{subfig:res1}--\ref{subfig:res2} for $c\in\{1,2\}$.
The curves resemble each other reasonably well, with $\qqSam$ generally 
predicting a slightly higher fingerprint quality than $\qqPo$ due to 
the implied independence assumptions in the latter. As $c$ increases, 
$\qqPo$ slowly approaches $\qqMLSame$. This indicates that the 
analytically derived $\qqPo$ is a good approximation of Mallory's  
fingerprint quality under idealized conditions particularly in 
high-frequency DCT subbands.

\paragraph{Different image sets}
In a more realistic scenario, we assume Alice and Mallory to work on 
different image sets.
Figures~\ref{subfig:res1} and \ref{subfig:res2} thus include $\phi_2$, 
the corresponding sample correlation coefficient between Alice's and 
Mallory's fingerprints as obtained with the ML estimator, averaged over 
five randomly compiled JPEG image sets of size $N_E=250$ that Alice has 
not used for computing $\bm{\hat{K}}$.
Alice's camera-specific fingerprint from 300 images was kept constant 
throughout all repetitions.
Not surprisingly, the population correlation 
coefficient $\qqPo$ is a loose upper bound to the observed correlation 
$\qqMLDiff$ when Mallory operates on a different image set: $\qqMLDiff$ approaches zero quickly with increasing $c$ and decreasing JPEG quality.
Appendix~\ref{sec:AppDifferentImageSets} gives additional 
insights by reporting the correlations for a much larger number of 
JPEG images, $N_E$. Mallory's fingerprint quality increases only slowly 
with the number of available JPEG images. For a suitable combination of 
JPEG quality and cut-off parameter~$c$, the correlation remains 
extremely small. As a result, less restrictive quality and cut-off 
parameters are possible compared to the contrived situation where Alice 
and Mallory access the same images.

\paragraph{Analysis summary}
Overall, strong guarantees for a scenario where Mallory has access to 
JPEG-compressed versions of the very images Alice used for her 
fingerprint are possible for JPEG qualities of 70 or smaller. In a more 
realistic scenario with different image sets, a secure operation is 
already possible with JPEG quality factors 90 or lower. For 
JPEG~85---the average quality factor on various image platforms (see 
Section~\ref{subsec:dataset})---Alice may choose $c \geqslant 3$ to 
ensure a reliable identification in an adversarial environment.

\subsection{(Q2) Independence Test}

The previous section has analyzed the correlation between Alice's and 
Mallory's high-frequency fingerprint estimates---deriving first bounds 
when Alice's fingerprint remains private.
We continue with this analysis under the scenario of different image sets
in the following sections.

Quiring and Kirchner have shown \cite{QuiringKirchner2015} 
that the high-frequency pixel part kept by $\bm{H}_c$ is uncorrelated 
to the complementary low-frequency part kept by \mbox{$\bm{L}_c = \bm{H}_c\ 
	\mathrm{XOR}\ 1$}. %
Consequently, Mallory cannot exploit linear dependencies between her 
robust low-frequency fingerprint and Alice's fragile fingerprint.
However, correlation does not cover all 
modes of dependence. In this section, we thus examine if Mallory can 
exploit non-linear dependencies and conduct a 
kernel statistical test of independence. In particular, we choose the 
Hilbert-Schmidt independence criterion\footnote{
	We use a Gaussian kernel, an alpha value %
	of 0.05, and split the 
	images into $320\times 320$ pixel blocks with varying offsets to 
	keep the sample size manageable.}
(HSIC)\cite{GreFukTeo+08}.
In simplified terms, 
this test maps the possibly non-linear dependencies to a linear 
space where independence is tested. The test is consistent in the sense 
that the level of $alpha$ controls the type I error (detects dependence 
although independence is true) while the type II errors goes to zero 
for an increasing sample size~\cite{GreFukTeo+08}. 

We consider the following two scenarios. First, we test if Alice's high 
frequency fingerprint is independent to Mallory's high-frequency 
fingerprint from JPEG-compressed images:
\begin{align}
\mathcal{H}_0 &: H_c(\bm{\hat{K}}) \independent H_c(\bm{\hat{K}_E}) \, .
\\\intertext{Equivalently, the second scenario tests if Alice's high 
frequency fingerprint is independent to Mallory's full fingerprint from 
JPEG-compressed images:}
\mathcal{H}_0 &: H_c(\bm{\hat{K}}) \independent \bm{\hat{K}_E} \, .
\end{align}
We grant Mallory access to $N_E=150$ images of each DDB camera and 
$N_E=1000$ RAISE images. %
We aggregate results over ten randomly compiled~sets of~size~$N_E$.

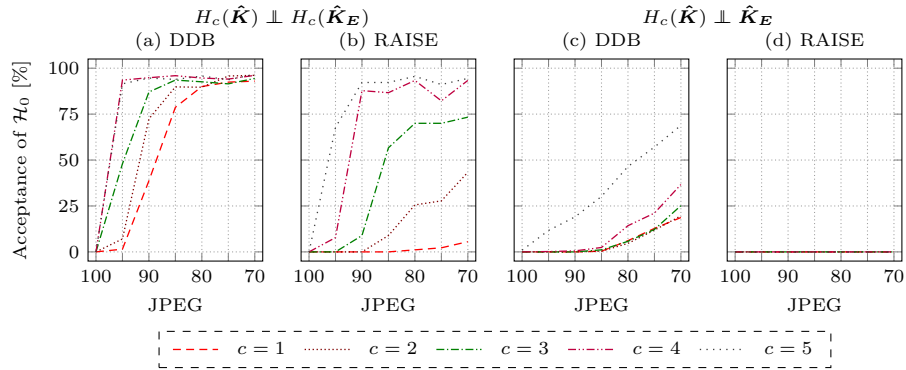
\begin{figure}[t]
	\centering
	\tikzsetfigurename{IndependenceFig}

\begin{tikzpicture}[font=\scriptsize] 

\def 
\pathtoindependenceA{./ext/Independence/independence_results_HF_HF_DRESDEN_permutation_150_.dat};
\def 
\pathtoindependenceB{./ext/Independence/independence_results_HF_HF_RAISE_permutation_1000_.dat};
\def 
\pathtoindependenceC{./ext/Independence/independence_results_HF_full_DRESDEN_permutation_150_.dat};
\def 
\pathtoindependenceD{./ext/Independence/independence_results_HF_full_RAISE_permutation_1000_.dat};

\def\titleA{$H_c(\bm{\hat{K}}) \independent H_c(\bm{\hat{K}_E})$};
\def\titleB{$H_c(\bm{\hat{K}}) \independent \bm{\hat{K}_E}$};

\begin{groupplot}
	[
	group style={group size=4 by 1,horizontal sep=0.51cm},
		width=.32\textwidth,
		height=.35\textwidth,
		xlabel={JPEG},
		ymax=1,
		ymin=0,
		ylabel near ticks,
		xlabel near ticks,
		xtick={1,0.95,0.9,0.85,0.8,0.75,0.7},
		xticklabels={100,,90,,80,,70},
		x dir=reverse,
		ytick={0,0.25,0.5,0.75,1},
		yticklabels={0,25,50,75,100},
		major tick length=3pt,
		enlarge x limits=0.05,
		enlarge y limits=0.05,
		clip mode=individual,
			major grid style={draw=gray!75,densely dotted},
						grid=both,
	]

\nextgroupplot[
		ylabel={Acceptance of $\mathcal{H}_0$ [\%]},
]		
	\addplot
	[
		no markers,
		line width=.5pt,
		coline2
	] table [x=JPEGS ,y =0] 
	{\pathtoindependenceA};
	
	\addplot
	[
		no markers,
		line width=.5pt,
		coline3
	] table [x= JPEGS ,y =1] 
	{\pathtoindependenceA};

	\addplot
	[
		no markers,
		line width=.5pt,
		coline4
	] table [x= JPEGS ,y =2] 
	{\pathtoindependenceA};
	
	\addplot
	[
		no markers,
		line width=.5pt,
		coline5
	] table [x= JPEGS ,y =3] 
	{\pathtoindependenceA};
	
	\addplot
	[
	no markers,
	line width=.5pt,
	coline6
	] table [x= JPEGS ,y =4] 
	{\pathtoindependenceA};

\node[font=\scriptsize,anchor=south] at (rel axis cs:0.5,1) {(a) 
DDB};

\nextgroupplot[
yticklabels={,,},
]		

\addplot
[
no markers,
line width=.5pt,
coline2
] table [x=JPEGS ,y =0] 
{\pathtoindependenceB};

\addplot
[
no markers,
line width=.5pt,
coline3
] table [x= JPEGS ,y =1] 
{\pathtoindependenceB};

\addplot
[
no markers,
line width=.5pt,
coline4
] table [x= JPEGS ,y =2] 
{\pathtoindependenceB};

\addplot
[
no markers,
line width=.5pt,
coline5
] table [x= JPEGS ,y =3] 
{\pathtoindependenceB};

\addplot
[
no markers,
line width=.5pt,
coline6
] table [x= JPEGS ,y =4] 
{\pathtoindependenceB};

\node[font=\scriptsize,anchor=south] at (rel axis cs:0.5,1) {(b) RAISE};

\nextgroupplot[
yticklabels={,,},
]		

\addplot
[
no markers,
line width=.5pt,
coline2
] table [x=JPEGS ,y =0] 
{\pathtoindependenceC};

\addplot
[
no markers,
line width=.5pt,
coline3
] table [x= JPEGS ,y =1] 
{\pathtoindependenceC};

\addplot
[
no markers,
line width=.5pt,
coline4
] table [x= JPEGS ,y =2] 
{\pathtoindependenceC};

\addplot
[
no markers,
line width=.5pt,
coline5
] table [x= JPEGS ,y =3] 
{\pathtoindependenceC};

\addplot
[
no markers,
line width=.5pt,
coline6
] table [x= JPEGS ,y =4] 
{\pathtoindependenceC};

\node[font=\scriptsize,anchor=south] at (rel axis cs:0.5,1) {(c) 
DDB};

\nextgroupplot[
yticklabels={,,},
		legend to name={CommonLegend},
		legend style={legend columns=5, column sep=5pt, dashed, inner 
		xsep=5pt}
]		

\addplot
[
no markers,
line width=.5pt,
coline2
] table [x=JPEGS ,y =0] 
{\pathtoindependenceD};
\addlegendentry{$c=1$}

\addplot
[
no markers,
line width=.5pt,
coline3
] table [x= JPEGS ,y =1] 
{\pathtoindependenceD};
\addlegendentry{$c=2$}

\addplot
[
no markers,
line width=.5pt,
coline4
] table [x= JPEGS ,y =2] 
{\pathtoindependenceD};
\addlegendentry{$c=3$}

\addplot
[
no markers,
line width=.5pt,
coline5
] table [x= JPEGS ,y =3] 
{\pathtoindependenceD};
\addlegendentry{$c=4$}

\addplot
[
no markers,
line width=.5pt,
coline6
] table [x= JPEGS ,y =4] 
{\pathtoindependenceD};
\addlegendentry{$c=5$}

\node[font=\scriptsize,anchor=south] at (rel axis cs:0.5,1) {(d) RAISE};
	
\end{groupplot}

\node[yshift=1.75em] at ($(group c1r1.north)!0.5!(group c2r1.north)$) 
{\titleA};
\node[yshift=1.75em] at ($(group c3r1.north)!0.5!(group c4r1.north)$) 
{\titleB};

\path (group c2r1.south east) -- 
node[below,yshift=-2.5em]{\ref{CommonLegend}} (group 
c3r1.south west);

\end{tikzpicture}
	\vspace{-2em}
	\caption{Kernel statistical test of independence. Plots (a) and (b) 
		depict the first scenario; Plots (c) and (d) 
		the second scenario for both databases.}
	\label{fig:IndependenceTest}
\end{figure}

For both scenarios, Figure~\ref{fig:IndependenceTest} depicts the 
observed
$\mathcal{H}_0$ acceptance rates, i.\,e. the percentage of cases for 
which we cannot detect a measurable dependence between the two quantities under test.
This rate increases with lower JPEG qualities or larger cut-off 
parameters $c$.
In the first scenario, the test statistic suggests independence
at a considerable rate for $c=4$ and JPEG quality 90  for all cameras.
Interestingly, the second scenario---where Alice's high 
frequency fingerprint is tested against Mallory's full 
fingerprint---is characterized by a lower rate of independence. 
A comparison of both scenarios thus suggests that remaining 
dependencies may result from the low-frequency part. This
dependency would have to be non-linear, as the low- and high-frequency 
signal are not correlated to each other. As we show in the next 
section, it is unclear how Mallory can exploit these potentially 
remaining dependencies in practice, however.

We surmise that the notably less conclusive results on the RAISE data 
are due to non-trivial remnants of image content in the noise 
residuals. In contrast to the Dresden database, Alice's fingerprint is 
here calculated from natural instead of homogeneously lit images, 
raising the bar for establishing independence considerably.

\paragraph{Analysis summary}
The chosen HSIC test establishes statistical independence for suitable JPEG and cut-off 
parameters, which gives a strong evidence that Mallory cannot exploit 
any dependence to recover Alice's fingerprint. Considering the high-frequency signals, Alice may choose $c\geqslant3$ for quality factor 85.

\subsection{(Q2) DCT Recovery}\label{subsec:eval_dctrecovery}

In the next experiment, we examine if Mallory can exploit 
remaining dependencies to recover DCT coefficients. Although a DCT 
coefficient that is quantized to zero does
not reveal information about the fingerprint, non-zero 
coefficients may leak information at least with their sign. By 
averaging enough images, Mallory may thus obtain a coarse fingerprint 
estimate. We test below if Mallory can improve her fingerprint by 
recovering DCT coefficients that were quantized to~zero.

We adapt the systematic approach by Li et al.~\cite{Li:2011}, since it is in principle also applicable to the recovery of
high-frequency DCT coefficients.
The recovery is a linear optimization problem with the objective to 
minimize the spatial distance of neighboring pixels within and across 
the $8\times 8$ pixel blocks from JPEG compression. The first 
constraint is that the recovered pixel values must correspond to their 
DCT coefficients. Second, DCT coefficients that should not be 
recovered are fixed. Finally, the pixel and DCT coefficients have to 
be within their dynamic range. The optimization problem can be 
summarized as
\begin{alignat}{2}
\min && & \textstyle \sum_{l,l'} \vert \x(l) - \x(l') \vert \\
\text{s.t. } && &\x -  \DCT^{\top} \ma \y \ma \DCT = \bm{0} \; , 
\label{eq:contraint1} \\
&& &\y(s) = \y^{*}(s) \; , \\
&& &\x(l) \in [x_{\min},x_{\max}] , \; \y(s) \in [y_{\min},y_{\max}] ,
\end{alignat}
where $l$ and $l'$ are the indices of neighboring pixels in 
the spatial domain, \DCT~denotes the DCT transformation matrix, and $s$ 
is the index of a DCT subband. The second constraint fixes with 
$\y^{*}$ all DCT coefficients 
that are not part of the subbands retained by filter $H_c$ or are 
non-zero in the subbands retained by~$H_c$. As a result, we recover 
only zero-valued DCT coefficients that $H_c$ retains.
For each image, and for each $8 \times \; 8$ pixel 
block, we set up such an optimization problem and include its 
direct neighboring blocks.

\begin{table}[tb]
		\setlength{\tabcolsep}{7pt}
		\scriptsize
	\centering
	\caption{Contingency table  of DCT recovery from 50~Nikon 
	D70 images}
	\label{tab:dctrecoveryconfusionmatrix}
	\begin{tabular}{crrrr|rrr}
		\toprule
		& & \multicolumn{6}{c}{Fraction Predicted Sign} \\
		& & \multicolumn{3}{c}{JPEG 100} & \multicolumn{3}{c}{JPEG 95} 
		\\
		\cmidrule(lr){3-5} \cmidrule(lr){6-8}
		& & neg & zero & pos &  neg & zero & pos   \\ 
		\midrule
		\multirow{3}{*}{\parbox{1.50cm}{Fraction\\True Sign}}  
		& neg & 0.08 & 0.09 & 0.05 & 0.11 & 0.15 & 0.09 \\ 
		& zero & 0.16 & 0.08 & 0.16 & 0.07 & 0.11 & 0.07 \\ 
		& pos & 0.05 & 0.09 & 0.24 & 0.09 & 0.15 & 0.17 \\ 
		\bottomrule
	\end{tabular} 
\end{table}

We report results for 50~images from a Nikon~D70 over the JPEG 
qualities 100 and 95 as well as the cut-off frequency $c=1$. The 
performance does not change considerably for smaller JPEG qualities or 
larger cut-off frequencies and thus are omitted.
Table~\ref{tab:dctrecoveryconfusionmatrix} depicts a contingency table 
that summarizes the frequency of correctly predicted signs. This is the 
case when the sign of the predicted DCT coefficient equals 
the sign from the corresponding original uncompressed image or both the 
predicted and uncompressed coefficient lie in the zero range 
$[-0.25,0.25]$.
Even for JPEG quality 100, the recovery cannot reliably predict the 
sign. The correct distinction drops further for a smaller JPEG 
quality and tends towards a random classifier.
In each case, the correlation to Alice's fragile fingerprint decreases 
when Mallory uses the recovered images for her 
estimate. %
In contrast, the recovery of low-frequency subbands is successful with 
an average recovery rate of 70\%. However, only the 
correlation to Alice's low-frequency fingerprint increases in our 
experiments.

\paragraph{Analysis summary}
The recovery of the correct sign is partly possible for 
low-frequency subbands, while the recovery of high-frequency 
subbands is already difficult for JPEG quality 100. Mallory can 
thus not improve her estimate of Alice's fragile fingerprint 
through a DCT recovery.

\subsection{(Q3) Fingerprint-Copy Attack} 
\label{subsec:eval_fingerprint-copy}

\begin{figure}[tb]
	\centering
\includegraphics[]{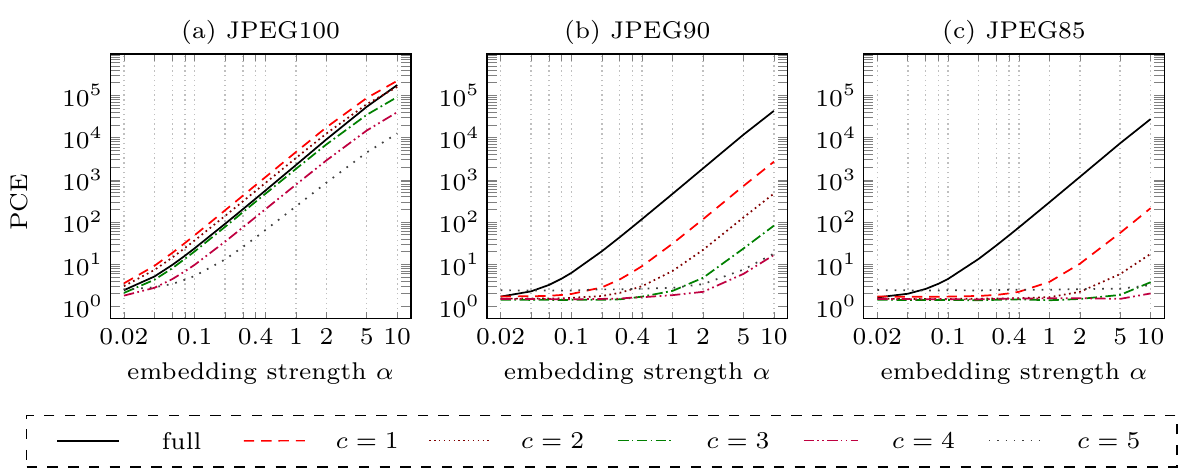}
	\vspace{-1.5em}
	\caption{Fingerprint-copy attack with fragile fingerprints. Average 
		PCE values as a function of the embedding strength $\alpha$ 
		with $N_E = 4648$ (Nikon D7000) for different JPEG qualities.
	}
	\label{fig:PCEofAlphasFigure}
\end{figure}

We finally consider a realistic fingerprint-copy attack 
where Mallory plants her calculated
fingerprint estimate $\hat{\prnu}_E$ from Alice's camera into 100 
randomly~chosen uncompressed images taken by a different camera (see 
Section~\ref{subsec:fingerprint-copy-method}).
Figure~\ref{fig:PCEofAlphasFigure} 
depicts the average PCE values with respect to the embedding strength 
for varying JPEG qualities. We present results only for the Nikon D7000 
from the RAISE database with $N_E = 4648$. This allows us to depict the 
effect when Mallory uses a large number of public images. We refer to 
Quiring and Kirchner~\cite{QuiringKirchner2015} for results from the 
DDB, which are similar to the results reported  here. 

As expected, high-quality JPEG~100 images enable Mallory to perform a 
successful attack due to the negligible quantization
(Fig.~\ref{fig:PCEofAlphasFigure}(a)).  
The situation is substantially different with stronger 
compression. For JPEG~90, only the full fingerprint gives high 
PCE values for small embedding strengths. Yet, larger cut-offs
demand extremely strong embeddings to achieve high PCEs.
For JPEG quality 85, no choice of $\alpha$ will produce Mallory's 
desired result with~$c\geqslant3$.

\paragraph{Analysis summary}
Fragile fingerprints allow a secure identification 
starting from JPEG~90 and lower.
In accordance to our results from previous sections,
no choice of $\alpha$ will allow an attack with 
	$c\geqslant3$ for quality factor 85.

\section{Application Analysis} \label{sec:Application}

We finally examine the application of fragile sensor noise fingerprints.
First, we verify that they are still discriminative enough to 
distinguish different cameras. Second, we compare them with the 
triangle test against fingerprint-copy attacks.

\subsection{Camera Identification}\label{subsec:eval_CamIdent}

\begin{figure}[t]
	\centering
	\captionsetup[subfigure]{labelformat=empty}
	\hspace{2.5em}
	\subfloat[]{\label{subfig:Application1} 
\includegraphics[]{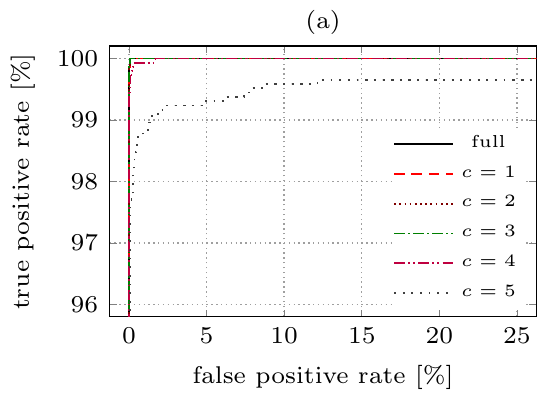}
	}
	\subfloat[]{\label{subfig:Application2}
\includegraphics[]{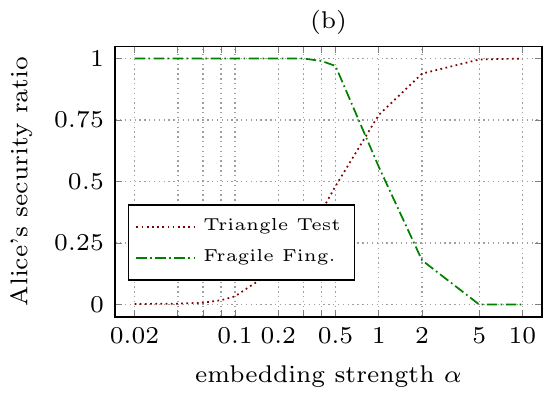}
	}
	\vspace{-1.5em}
	\caption{Applications. Plot (a) shows the camera identification of 
	uncompressed images. Plot (b) depicts the defense 
	performance against fingerprint-copy attack using the triangle test 
	and fragile fingerprints (JPEG quality 90, $c=1$).}
	\label{fig:Application1x2}
\end{figure}

In the following, we show 
that fragile fingerprints allow a reliable camera identification 
compared to traditional \emph{full}~camera fingerprints.
We only consider uncompressed images here by the very nature of fragile 
fingerprints. The PCE is used as similarity measure for images 
of each camera (true positives) and all 
remaining natural images from the Dresden Image Database (true 
negatives).
Figure~\ref{subfig:Application1} shows the ROC 
curves for different cut-off frequencies~$c$---aggregated over 1442 
images from the six DDB cameras. The full frequency 
range is included for comparison.

Although a fragile fingerprint with $c=1$ employs only 28~DCT 
coefficients in each block, it achieves the same detection performance 
as the full fingerprint with 64~coefficients. An almost perfect 
detection is possible with $c\leqslant4$ for the Dresden database.
The results for the Nikon D7000 camera are comparable for
$c\leqslant3$. We contribute this smaller choice of $c$ to a more 
perturbed fingerprint estimate of this camera---due to more 
image content in the respective noise residuals.

\paragraph{Analysis summary}
Fragile fingerprints allow a reliable camera identification. 
Together with our security analysis, for a common JPEG quality 
factor of 85, Alice can choose $c=3$ to achieve both a reliable camera 
identification and attack resistance.

\subsection{Comparison with Triangle 
Test}\label{subsec:eval_triangletest}

While the triangle test cannot be recommended for authentication, it is 
a reasonable defense in digital image forensics. 
Our final experiment highlights its powerful combination with fragile 
fingerprints in forensic applications against fingerprint-copy attacks.
Our previous results underline that remaining fingerprint information 
after quantization are usable for large embedding strengths with 
too small cut-off parameters ($c\leqslant 2$). However, the 
triangle test shows its strengths exactly in these cases, as the 
additional residual image content from the forgery process emerges more 
clearly with a larger embedding strength~\cite{Goljan:2011aa}.

In the following, we assume Mallory to have access to $N_E=150$ public JPEG 
images with quality factor 90 from Alice's Nikon D200 camera.
Mallory embeds her spoofed fingerprint into an 
uncompressed image from another camera while varying $\alpha$ as 
defined in the previous section. On the defender side, the linear 
parameters $\theta$ and $\mu$ of the test statistic are 
estimated from 200~images that Mallory has not used, cf. 
Eq.~\eqref{eq:triangle-test}. We set the threshold $t$ such that 
the false alarm probability is $10^{-3}$. Finally, Alice reports her 
security ratio: the percentage of images that Mallory has used and that 
are correctly marked as those. We repeat the process over 100 randomly 
chosen uncompressed images where Mallory embeds her spoofed 
fingerprint. Figure~\ref{subfig:Application2} depicts the averaged 
security ratio.

For reference, we include the corresponding results with the fragile 
fingerprint approach. We focus on $c=1$ and JPEG quality 90, 
where Mallory obtains considerably high PCE values with a
fingerprint-copy attack
(see~Figure~\ref{fig:PCEofAlphasFigure}(b)). In particular, we estimate 
the distribution of the PCE values from uncompressed images with a 
Gaussian kernel density estimator. The PCE threshold under which an 
image is not assumed as one from Alice's camera is set such that the 
false positive probability is $10^{-3}$. Alice's security ratio 
expresses the percentage of Mallory's images that do not exceed the PCE 
threshold and thus are correctly identified as being not from Alice's 
camera.

Figure~\ref{subfig:Application2} emphasizes that both 
approaches are a powerful combination when Mallory has just access to 
JPEG images. 
At the point where Mallory starts to circumvent fragile 
fingerprints, the triangle test already detects more than 50\% of 
images that are involved in Mallory's attack; usually enough to raise 
suspicion that Mallory has forged the image under investigation. In 
summary, Mallory faces the following dilemma: A too strong fingerprint 
strength is likely to be uncovered by the triangle test; with a too 
weak embedding, Mallory's forged image will not be identified as one of 
Alice's images. By using the triangle test in addition, Alice can even 
use smaller cut-off values for her fragile fingerprint.

\section{Conclusion} \label{sec:conclusion}

This paper contributes to a thorough understanding of fragile 
camera fingerprints by providing a comprehensive security analysis.
In multiple tests, we confirm that Mallory cannot estimate Alice's 
camera fingerprint from JPEG-compressed images with common compression 
levels. Our analysis thus motivates the usage of fragile fingerprints 
in various applications, such as authentication or digital image 
forensics.
Finally, we note that the concept of fragile fingerprints effectively 
demonstrates how asymmetries in the quality of accessible data can be 
exploited. In the context of recent unification attempts 
between related research 
disciplines~\cite{QuiArpRie18,Barni:2013aa}, this 
may foster novel strategies in adversarial machine learning or signal 
processing.

\section*{Acknowledgments}
The authors gratefully acknowledge funding from Deutsche 
Forschungsgemeinschaft (DFG) under the project \mbox{RI 2469/3-1}, 
from the German Federal Ministry of Education and Research (BMBF) 
under the project FIDI (FKZ 16KIS0786K), and by the NSF grant 1464275. 
The first author also thanks the German Academic Exchange Service 
(DAAD) for financial support during his stay in Binghamton.

\appendix

\section{Sample Correlation Coefficient}\label{sec:AppCorrInDCT}

The objective is to compute Pearson's sample correlation coefficient 
between two images $\bm{u}$ and $\bm{v}$ equivalently in the DCT 
domain. Without loss of generality, we focus on an $8 \times 8$ pixel 
block, so that the correlation is given as 
\begin{align}
r = \dfrac{
	n \sum u(l) v(l) - \sum u(l) \sum v(l)
}{ n \;
\sqrt{\sum u(l)^2 - \left( \sum u(l)\right)^2}
\sqrt{\sum v(l)^2 - \left( \sum v(l)\right)^2}
} \label{eq:appendix_corr}
\end{align}
where $u(l)$ and $v(l)$ are the pixel values.
The total number of pixel values or DCT coefficients is given by $n$, 
thus for one block $n=64$.
To obtain the same correlation value just with the DCT representation 
$\bm{U}$ and $\bm{V}$ of both images, we use the following 
identities between spatial pixels and DCT~coefficients:
\begin{align}
n \sum u(l) v(l) &= n \sum U(l) V(l) \\
\sum u(l) \sum v(l) &= n^2 \; \bar{\bm{u}} \; \bar{\bm{v}} = n \; U(0) 
V(0) 
\\
n \sum u(l)^2 &= n \sum U(l)^2 \\
\left( \sum u(l) \right)^2 &= (n \; \bar{\bm{u}})^2 = n \; U(0)^2 \; .
\end{align}
Incorporating these identities in Eq.~\eqref{eq:appendix_corr} and 
canceling $n$, we obtain the following correlation equation for one $8 
\times 8$ pixel block:
\begin{align}
r = \dfrac{\sum \bm{U}(l) \bm{V}(l) - \bm{U}(0) \bm{V}(0)}
{\sqrt{\strut \sum \left( \bm{U}(l)^2 \right) - \bm{U}(0)^2}
	\sqrt{\strut \sum \left( \bm{V}(l)^2 \right) - \bm{V}(0)^2}
} \label{eq:appendix_corr_dct}
\end{align}
The generalization over all image blocks yields the 
same result. If we now just 
focus on AC coefficients, the DC coefficient $U(0)$ and $V(0)$ become 
zero. As the AC coefficient's mean goes to zero, 
Eq.~\eqref{eq:appendix_corr_dct} corresponds to 
Eq.~\eqref{eq:appendix_corr}. In other words, we can directly feed the 
AC DCT coefficients into the standard Pearson correlation equation. 

\section{Population Correlation Coefficient}
\label{sec:AppQuantization}
Given a quantizer and uniform step size $q$, we denote by $U$ an 
uncompressed image as random variable and by $V$ its quantized output, 
$V = \lfloor U/q + 0.5\rfloor\cdot q$. The objective 
is to compute the population correlation coefficient between $U$ 
and~$V$:
\begin{align}
\rho= \dfrac{\mathrm{Cov}(U, 
	V)}{\sqrt{\mathrm{Var}(U)}\sqrt{\mathrm{Var}(V)}} \; .
\end{align}
The following general relations between the random variable $U$ and 
output $V$ can be established when $U$ is assumed to have a symmetrical 
and zero-mean pdf $f_U(x)$ with characteristic function 
$\Phi_U(x)$~\cite{Widrow:2008}:
\begin{flalign}
\text{Var}(V) = \text{Var}(U) + \frac{q^2}{12} &+ \frac{q^2}{\pi^2} 
\sum_{k=1}^{\infty} \Phi_U \left(\frac{2\pi k}{q} \right) \cdot 
\frac{(-1)^k}{k^2} \nonumber & \\ &+
\frac{2q}{\pi} \sum_{k=1}^{\infty} \Phi^{'}_U \left(\frac{2\pi k}{q} 
\right) 
\frac{(-1)^{k+1}}{k} 
& \label{eq:QuantizationVarY} \\
\text{Cov}(U,V) = \text{Var}(U) &+ \frac{q}{\pi} \sum_{k=1}^{\infty} 
\Phi^{'}_U \left(\frac{2\pi k}{q} 
\right) \frac{(-1)^{k+1}}{k}  & \label{eq:QuantizationCov}
\end{flalign}
For a zero-mean Laplacian distribution 
with parameter $\lambda$, the characteristic function is given as:
\begin{align}
\Phi_U(x) = \frac{\lambda^2}{x^2 + \lambda^2} \;.
\end{align}
To highlight that the covariance and variance terms are only based on 
variable~$U$, we write $\text{Cov}(U,V) = \text{Cov}^+(U)$ 
and $\mathrm{Var}(V) = \mathrm{Var^+}(U)$.

In the next step, to determine the fingerprint quality, we need to 
calculate the correlation after averaging the uncompressed images and 
their compressed counterparts, respectively:
\begin{align}
\mathrm{cor}\left( \textstyle  \sum_i U_i , \sum_i V_i\right) 
\; . 
\label{eq:Appendix_RhoTheoModel}
\end{align}
We start with the distribution on one subband and denote 
by $U_{i,s}$ and $V_{i,s}$ the $s$-th subband of the $i$-th image and 
its compressed version. 
The aggregation over various subbands follows from the linear property of the covariance and the assumption of uncorrelated DCT subbands:
\begin{align}
\mathrm{Cov} \left( \textstyle \sum_s U_{i,s} , \sum_s V_{i,s} \right) &= \textstyle \sum_s 
\mathrm{Cov} (U_{i,s} , V_{i,s}) \; .
\label{eq:CovCarEq}
\end{align}
This is also possible for the variance of a sum of random variables.
Finally, we assume the images to be uncorrelated to average the covariance 
over all images:
\begin{align}
\mathrm{Cov} \left( \textstyle \sum_i U_{i} , \sum_i V_{i} \right) &= \textstyle \sum_i 
\sum_s \mathrm{Cov} (U_{i,s} , V_{i,s}) \; .
\label{eq:CovCarEq2}
\end{align}
Taking all together, the population correlation coefficient is given as
\begin{align}
&\rho\left( {\textstyle\sum_i U_i , \sum_i V_i }\right) %
= \dfrac{\sum_i \sum_s 
	\mathrm{Cov^+}(U_{i,s})}{\sqrt{\strut\sum_i\sum_s 
		\mathrm{Var}(U_{i,s})}\sqrt{\strut\sum_i 
		\sum_s \mathrm{Var^+}(U_{i,s})}} \; . 
		\label{eq:Appendix_PopCorCoeff}
\end{align}

\section{Empirical Quality of Fingerprint Estimation
}\label{sec:AppDifferentImageSets}
For the different image set scenario, 
Table~\ref{tab:tabattackFourLargeRAISE} shows the correlations for 
$N_E=2000$ and $N_E=4648$ JPEG images. $N_E=4648$ is the maximum number 
of available images in our setup, so that its values 
are from a single instance of the experiment. %

\vspace{-.15cm}
\begin{table}
	\setlength{\tabcolsep}{6.350pt}
	\scriptsize
	\caption{Quality of fingerprint estimation (RAISE)}
	\label{tab:tabattackFourLargeRAISE}
	\centering
	\begin{tabular}{cccccccc}
		\toprule
		\multirow{2}{*}{$N_E$}  & \multirow{2}{*}{JPEG}  & 
		\multicolumn{6}{c}{$c$} \\ 
		\cmidrule{3-8}
		&  & full & 1  & 2 & 3 & 4 & 5  \\ 
		\midrule 
		\multirow{7}{*}{2000}
		& 100 & \phantom{-}0.6241 & \phantom{-}0.6070 & 
		\phantom{-}0.5613 & 
		\phantom{-}0.4867 & \phantom{-}0.3858 & \phantom{-}0.2654  \\ 
		& 95 & \phantom{-}0.5484 & \phantom{-}0.3853 & 
		\phantom{-}0.2694 & 
		\phantom{-}0.1622 & \phantom{-}0.0824 & \phantom{-}0.0551  \\ 
		& 90 & \phantom{-}0.4633 & \phantom{-}0.1588 & 
		\phantom{-}0.0793 & 
		\phantom{-}0.0375 & \phantom{-}0.0181 & \phantom{-}0.0185  \\ 
		& 85 & \phantom{-}0.4030 & \phantom{-}0.0550 & 
		\phantom{-}0.0178 & 
		\phantom{-}0.0071 & \phantom{-}0.0031 & \phantom{-}0.0065  \\ 
		& 80 & \phantom{-}0.3619 & \phantom{-}0.0195 & 
		\phantom{-}0.0026 & 
		-0.0003 & \phantom{-}0.0002 & \phantom{-}0.0045  \\ 
		& 75 & \phantom{-}0.3301 & \phantom{-}0.0070 & -0.0029 & 
		-0.0024 & 
		\phantom{-}0.0004 & \phantom{-}0.0023  \\ 
		& 70 & \phantom{-}0.3093 & \phantom{-}0.0035 & -0.0042 & 
		-0.0036 & 
		-0.0001 & \phantom{-}0.0017 \\ 
		\midrule
		\multirow{7}{*}{4648}
		& 100 & \phantom{-}0.6526 & \phantom{-}0.6387 & 
		\phantom{-}0.5968 & 
		\phantom{-}0.5279 & \phantom{-}0.4330 & \phantom{-}0.3151  \\ 
		& 95 & \phantom{-}0.5890 & \phantom{-}0.4654 & 
		\phantom{-}0.3531 & 
		\phantom{-}0.2269 & \phantom{-}0.1208 & \phantom{-}0.0808  \\ 
		& 90 & \phantom{-}0.5054 & \phantom{-}0.2162 & 
		\phantom{-}0.1130 & 
		\phantom{-}0.0543 & \phantom{-}0.0264 & \phantom{-}0.0277  \\ 
		& 85 & \phantom{-}0.4451 & \phantom{-}0.0757 & 
		\phantom{-}0.0248 & 
		\phantom{-}0.0099 & \phantom{-}0.0040 & \phantom{-}0.0094  \\ 
		& 80 & \phantom{-}0.4045 & \phantom{-}0.0272 & 
		\phantom{-}0.0033 & 
		-0.0008 & -0.0005 & \phantom{-}0.0047  \\ 
		& 75 & \phantom{-}0.3732 & \phantom{-}0.0100 & -0.0044 & 
		-0.0035 & 
		\phantom{-}0.0006 & \phantom{-}0.0033  \\ 
		& 70 & \phantom{-}0.3535 & \phantom{-}0.0053 & -0.0058 & 
		-0.0053 & 
		-0.0003 & \phantom{-}0.0022  \\ 
		\bottomrule 
	\end{tabular} 
\end{table}

\end{document}